\documentclass[traditabstract]{aa} 
\usepackage{longtable}
\usepackage{txfonts,epsfig,graphicx,natbib,url,twoopt}
\usepackage[breaklinks=true]{hyperref} 
\usepackage[labelfont=footnotesize,textfont=footnotesize]{caption}

\usepackage{verbatim}
\usepackage{natbib}
\bibpunct{(}{)}{;}{a}{}{,} 

\newcommand{\kms}{km~\hspace{0.5mm}s$^{-1}$ \ }
\newcommand{\vs}{$v$~sin~$i$}
\usepackage[utf8]{inputenc}

\begin{document}

\title{A catalog of rotational and radial velocities for evolved stars V. Southern stars
\thanks{Based on observations collected at the Haute--Provence Observatory, Saint--Michel, France,
and at the European Southern Observatory, La Silla, Chile}\fnmsep\thanks{
   Table 1 is only available in electronic form at the CDS via anonymous
   ftp to {\tt cdsarc.u-strasbg.fr} (130.79.128.5) or via
   {\tt http://cdsweb.u-strasbg.fr/cgi-bin/qcat?J/A+A/???/???}} }

   \subtitle{}

   \author{J. R. De Medeiros\inst{1},
                 S. Alves\inst{1},    
             S. Udry\inst{2}, 
		J. Andersen\inst{3,4},
		B. Nordstr{\"o}m\inst{3} and
                 M. Mayor\inst{2}}

   \offprints{J. R. De Medeiros}

\institute{Departamento de F\'isica, Universidade Federal do Rio Grande do Norte, Campus Universit\'ario, 59072-970 Natal, RN, Brasil\\
     \email{J. R. De Medeiros, renan@dfte.ufrn.br}
\and
  Observatoire de Gen\`eve, Universit\`e de Gen\`eve, Chemin des Maillettes 51, 1290 Sauverny, Switzerland
\and
The Niels Bohr Institute, University of Copenhagen, Juliane Maries Vej 30, 2100 Copenhagen, Denmark
\and
Nordic Optical Telescope, Apartado 474, 38700 Santa Cruz de La Palma, Spain 
}

\authorrunning {J.~R. \ De Medeiros et al.}
\titlerunning{A catalog of rotational and radial velocities for evolved stars}
\date{Received Date:?? Accepted Date:??}

  \abstract{ 
Rotational and radial velocities have been measured for 1589 evolved stars of spectral types F, G, and K and luminosity classes IV, III, II, and Ib, based on observations carried out with the CORAVEL spectrometers. 
 The precision in radial velocity is better than 0.30 km\hspace{0.5mm}s$^{-1}$ per observation, whereas rotational velocity uncertainties are typically 1.0 \kms for subgiants and giants and 2.0 km\hspace{0.5mm}s$^{-1}$ for class II giants and Ib supergiants.
}

\maketitle

\section{Introduction}\label{intro}

Over the past two decades, observations have been carried out at the Geneva Observatory, Switzerland, and the Federal University of Rio Grande do Norte, Brazil, to accurately measure projected rotational velocities (\vs) of evolved stars, with the aim of studying the evolution of stellar rotation with stellar age.

The technique used is to combine a high-resolution spectrometer with the cross--correlation technique, which yields accurate, high S/N cross--correlation line profiles from relatively low S/N spectra. From these profiles, accurate radial velocities and, once calibrated, projected rotational velocities (\vs) with an accuracy better than 1 km\hspace{0.5mm}s$^{-1}$ can be derived, allowing measurements of \vs\ for large samples of relatively faint stars with telescopes 
of moderate aperture. 

Most of the observations presented here were made with the CORAVEL cross--correlation spectrometers \citep{1979VA.....23..279B}. In addition, \citet{2006A&A...458..895D} measured rotational velocities, \vs\ , for 100 metal-poor stars with the digital version of the cross--correlation procedure, using spectra obtained with the FEROS \citep{1998SPIE.3355..844K} and CORALIE \citep{1996A&AS..119..373B} spectrometers.

As part of this programme, \citet{1999A&AS..139..433D} measured \vs\ for 1541 stars
of luminosity classes IV, III, and II, \citet{2002A&A...395...97D} presented \vs\ for 232 Ib supergiant stars, and \citet{2004A&A...427..313D} also measured \vs\ for 78 double--lined binaries with an evolved component. These high--quality data have inspired several studies, enabling reliable investigations of stellar rotational characteristics in different regions of the H--R Diagram (Carlberg et al. \citeyear{2011ApJ...732...39C}; \citeauthor{2009ApJ...704..750C} \citeyear{2009ApJ...704..750C}; \citeauthor{2001A&A...375..851M} \citeyear{2001A&A...375..851M}), the relationship between rotation and different stellar properties \citep{2011A&A...529A..90M,2010ApJS..190....1R,2010A&A...514A..97L,2010MNRAS.408.2290G,2008AJ....135..209M,2002ApJ...578..943D,2002A&A...384..491C}, constraints on theoretical models \citep{2010A&A...509A..72E,2009ApJ...702.1078B} and in many studies on extra-solar planets \citep[e.g.][]{2010MNRAS.408.1606W,2009A&A...505.1311D}.

The present work brings complementary results for our observational efforts, with the measurements of projected rotational velocity \vs\  for southern subgiant, giant, bright giant, and Ib supergiant stars of spectral types F, G, and K, listed in the Bright Star Catalog \citep{1982bscf.book.....H,1983stbs.book.....H}. Although the primary aim of this investigation is to study the rotational behaviour of evolved stars, our observational procedure also produced a large set of radial velocity measurements, representing an important tool for answering several questions in stellar astrophysics, including the search for planets around evolved stars. 

This paper is arranged as follows. Section 2 presents the definition of the sample, the observational procedure used throughout this survey, and the calibration of rotational velocities, with a discussion of their probable errors. The list of individual \vs\ measurements and mean radial velocities are presented in Sect. 3.

\section{The observational programme}

The present sample consists of a total of 1702 mainly southern F, G, and K stars of luminosity classes IV, III, II, and Ib listed in the Bright Star Catalog. 
Most of these stars were observed in different programmes carried out at the
Geneva Observatory, the majority devoted to studying stellar binarity \citep{1991A&AS...88..281D, 1999A&AS..139..433D,2002A&A...395...97D,2004A&A...427..313D}  or to precise measurements of radial velocity in programmes on Galactic structure \citep{1985A&AS...59...15A,1985A&AS...62...23P,1987A&AS...67..423M,2004A&A...418..989N}.

As in previous papers \citep{1999A&AS..139..433D,2002A&A...395...97D,2004A&A...427..313D}, the observations reported here were made using the two CORAVEL spectrometers \citep{1979VA.....23..279B} mounted on the 1.54-m Danish telescope at ESO, La Silla (Chile), and the 1-m Swiss telescope at Haute-Provence Observatory, Saint Michel (France). Radial velocities were derived by direct cross--correlation of the stellar spectra with a binary (0, 1) physical template, constructed from the spectrum of the K2 III star Arcturus and mounted inside the spectrometers. The radial-velocity system applied is that defined by \citet{1999ASPC..185..367U}. Typical integration times were 5 min, and data was reducted using standard procedures \citep{1987A&A...178..114D,1991A&AS...88..281D,1999A&AS..139..433D}. For a complete discussion of the observational procedure, calibration, and error analysis, readers are referred to \citet{1987A&A...178..114D}, \citet{1991A&AS...88..281D}, and \citet{1999A&AS..139..433D}. 

Here, we just recall a few salient points. In all cases, the radial velocity uncertainty is derived from an instrumental error added in quadrature to photon and scintillation noise, which are estimated using the computed parameters of the cross--correlation profiles \citep{1979VA.....23..279B}. Different studies of large data samples \citep{1991A&AS...88..281D,1997ESASP.402..693U,1999A&AS..139..433D} show that the typical uncertainty for  CORAVEL radial velocity is about 0.3~km\hspace{0.5mm}s$^{-1}$ for slowly rotating stars, generally with   \vs~$<$~20~km\hspace{0.5mm}s$^{-1}$. For faster rotators, the uncertainty is somewhat greater. 

Rotational velocities (\vs) were obtained through an appropriate calibration of the widths of cross--correlation profiles, as described by \citet{1999A&AS..139..433D}. The original \vs\ calibration by \citet{1984A&A...138..183B} is also valid for subgiant and giant stars of luminosity classes IV and III, but for class II and Ib bright giants and supergiants, the increase in macroturbulence with spectral type required a new calibration of the width of the cross--correlation profile into \vs\ as measured from a Fourier transform of line profiles from \citet{1986ApJ...310..277G,1987ApJ...322..360G}.
{Whereas  \citet{1984A&A...138..183B} obtained for the parameter associated with the CORAVEL cross--correlation profiles $\sigma_0$ the value of 6.88~km\hspace{0.5mm}s$^{-1}$ for stars of luminosity classes V to III, the new calibration shows that for the luminosity classes II and Ib the value of $\sigma_0$ is 7.158~km\hspace{0.5mm}s$^{-1}$ and 7.978~km\hspace{0.5mm}s$^{-1}$, respectively.}

The computed \vs\ has a typical uncertainty of around 1.0 km\hspace{0.5mm}s$^{-1}$ for subgiant and giant stars with \vs\ $< <$30~km\hspace{0.5mm}s$^{-1}$, whereas for bright giants and Ib supergiants, we conservatively assume an uncertainty of 2.0 km\hspace{0.5mm}s$^{-1}$, since it is impossible to define precise limits between rotation and macroturbulence. For faster rotators, those with \vs\  higher than 30~km\hspace{0.5mm}s$^{-1}$, \citet{1999A&AS..139..433D} estimate an uncertainty of about 10\%, regardless of luminosity class. 

\section{Contents}

The main results of this catalogue are listed in Table 1, which presents CORAVEL rotational and mean radial velocities for 1589 evolved FGK stars of luminosity classes IV, III, II, and Ib, ordered by HD number, for single stars and single-lined spectroscopic binaries. Columns are as follows:\\
\noindent
1. HD number;\\
2. Spectral type;\\
3. (B -- V) color index;\\ 
{ 4-5. Mean radial velocity RV and its uncertainty $\epsilon$, on $N$ number of CORAVEL observations. In this case, the uncertainty is given by max ($  \epsilon_1/\sqrt{N}$, $\sigma/\sqrt{N}$), where $\epsilon_1$ is the typical error for one single radial velocity measurement;}\\
6. Radial velocity dispersion (rms) $\sigma$;\\
7. E/T, the ratio of observed to expected rms dispersion for observations, when $N \geq 2$;\\
8. P($\chi^2$), the probability that the radial velocity of the star is constant;\\
9. $N$, number of observations for each star;\\
10. Time span $\Delta$ T of observations;\\
11-12. Rotational velocity $V \sin i$ and its uncertainty $\epsilon_{\textrm{rot}}$;\\
13. Remarks. The remarks SBO, SB, and SB? indicate, respectively, single-lined spectroscopic binaries for which orbital parameters are available in the literature, stars displaying single-lined spectroscopic binary behaviour, and stars for which the dispersion or any systematic trend in the CORAVEL velocities suggest that they may be single--lined binaries, but for which no period could be determined. 
Nevertheless, for a few stars classified as SB and SB?, the RV variability may reflect another cause, e.g. pulsation. 

A number of the programme stars, identified in Table 1, were already included in the papers by \citet{1985A&AS...59...15A}, \citet{1985A&AS...62...23P}, and \citet{1987A&AS...67..423M}, with radial velocities referred to in the 1985 standard system of \citet{1999ASPC..185..367U}, while the velocities listed here are referred to in the revised zero-point of \citet{1999ASPC..185..367U}. The colour-dependent differences from the earlier velocities are small, about 0.11, 0.31, and 0.45 km\hspace{0.5mm}s$^{-1}$ for F--, G--, and K--type stars, respectively, but can be noticed in precise work.

Table 2 lists 79 SB2 and SB3 binary systems also observed by CORAVEL, many of which were detected here for the first time, while Table 3 presents the evolved F--type stars for which no correlation dip was obtained with CORAVEL. These are undoubtedly fast rotators.

{The individual radial velocity measurements for single and SBO stars, as well as for SB not included in follow--up programmes, are available at the CDS ``Centre de Donn\'ees Astronomiques de Strasbourg''. }

\bibliographystyle{aa}
\begin{acknowledgements}
We gratefully acknowledge the support of the CORAVEL observers in this survey, particularly Gilbert Burki, Bernard Pernier, Pierre North, Jean--Claude Mermilliod and Gerard Jasniewicz. We also thank Emile Ischi and Bernard Tartarat for technical maintenance of the CORAVELs and B. L. Canto Martins, I. C. Le\~ao and J. D. do Nascimento for assistance in preparing tables and data controls. S. Alves acknowledges a graduate fellowship and a PNPD fellowship from the CAPES Brazilian agency. This study used the SIMBAD database, operated at the CDS in Strasbourg, France, and it was supported by continuous grants from the Swiss National Science Foundation. Research activities of the Observational Stellar Board of the Federal University of Rio Grande do Norte are supported by the Brazilian agencies CNPq, FAPERN, and the INCT INEspa\c{c}o. The observations at the Danish 1.54m telescope from ESO, La Silla, Chile, was supported by grants from ESO and Danish observing time, and financially by the Danish Natural Science Research Council through the Danish Board for Astronomical Research.
\end{acknowledgements}


\onecolumn


\tiny{
\begin{longtable}{|lcr|rrrrr|rr|rr|l|} 
\caption{Rotational and radial velocities for evolved stars}\\
\hline 
HD Number	&	Sp	&	(B - V)	&	RV	&	$\epsilon$	&	$\sigma$	&	E/T	&	P($\chi^2$)	&	N	&	$\Delta$ T	&	$V \sin i$	&	$\epsilon_{\textrm{rot}}$	&	Remark	\\
	&		&		&	km s$^{-1}$	&	km s$^{-1}$	&	km s$^{-1}$	&		&		&		&	days	&	km s$^{-1}$	&	km s$^{-1}$	&		\\
\hline
\endfirsthead
\caption{continued}\\
\hline 
HD Number	&	Sp	&	(B - V)	&	RV	&	$\epsilon$	&	$\sigma$	&	E/T	&	P($\chi^2$)	&	N	&	$\Delta$ T	&	$V \sin i$	&	$\epsilon_{\textrm{rot}}$	&	Remark	\\
	&		&		&	km s$^{-1}$	&	km s$^{-1}$	&	km s$^{-1}$	&		&		&		&	days	&	km s$^{-1}$	&	km s$^{-1}$	&		\\
\hline
\endhead
\hline
\endfoot
\multicolumn{13}{l}{{{Notes: $^\dagger$:  Radial velocity measurements were also determined by \cite{1985A&AS...59...15A}, using the radial velocity system by \cite{1985srv..conf..299M}.}}}\\
\multicolumn{13}{l}{{{The mean radial velocities given in the present work for a total of 456 stars in common with  \cite{1985A&AS...59...15A},  show a difference of about 0.11, }}}\\
\multicolumn{13}{l}{{{0.31 and 0.45 km\hspace{0.5mm}s$^{-1}$, respectively for F--, G-- and K--type stars in relation to  previous measurements. }}}\\
\multicolumn{13}{l}{{{{$^\ddagger$ Vsini values also listed by \cite{2004yCat.5117....0N}.}}}}\\
\hline
\endlastfoot
142$^\ddagger$	&	G1IV	&	0.52	&	5.45	&	0.28	&	0.10	&	0.25	&	0.803	&	2	&	1528	&		11.3	&	1.0	&		\\
344$^\dagger$	&	K1III	&	1.12	&	7.15	&	0.19	&	0.14	&	0.54	&	0.586	&	2	&	212	&	$<$	1.6	&		&		\\
469$^\ddagger$	&	G4IV	&	0.74	&	-0.90	&	0.23	&	0.17	&	0.53	&	0.593	&	2	&	1050	&		2.7	&	2.0	&		\\
496	&	K0III	&	1.03	&	-8.57	&	0.19	&	0.26	&	0.96	&	0.339	&	2	&	803	&	$<$	1.5	&		&		\\
587	&	K1III	&	0.98	&	24.39	&	0.19	&	0.13	&	0.48	&	0.631	&	2	&	1508	&	$<$	1.0	&		&		\\
636	&	G8III	&	1.05	&	13.71	&	0.61	&	0.87	&	3.19	&	0.001	&	2	&	1128	&	$<$	1.0	&		&		\\
720	&	K5III	&	1.34	&	-5.88	&	0.48	&	0.67	&	2.58	&	0.010	&	2	&	313	&		2.2	&	1.2	&		\\
905$^\ddagger$	&	F0IV	&	0.31	&	-25.71	&	1.30	&	1.83	&	1.30	&	0.199	&	2	&	387	&		31.6	&	5.8	&		\\
\end{longtable}
}


{
\begin{longtable}{r c c }
\caption{Double-lined spectroscopic binary systems SB2 with evolved component}\\
\hline 
HD	&	(B - V)	&	ST	\\
\hline
\endfirsthead
\caption{continued}\\
\hline 
HD	&	(B - V)	&	ST	\\
\hline
\endhead
\hline
\endfoot
352	&	1.38	&	K2III+F	\\
5357	&	0.38	&	 F4III	\\
5516	&	0.94	&	 G8IIIb	\\
6397	&	0.41	&	 F4II-III	\\
7788	&	0.47	&	 F6IV	\\
8949	&	1.12	&	 K1III	\\
10308	&	0.44	&	 F2III	\\
12641	&	0.88	&	 G5II-III+G5V	\\
\end{longtable}
}

{
\begin{longtable}{c c }
\caption{Evolved stars with no CORAVEL dip.}\\
\hline 
HD	&	ST	\\
\hline
\endfirsthead
\caption{continued}\\
\hline 
HD	&	ST	\\
\hline
\endhead
\hline
\endfoot
%
%
%
432	&	F2III-IV	\\
571	&	 F2II	\\
4338	&	 F1IVn	\\
6763	&	 F0III-IV	\\
6953	&	 K7III	\\
11522	&	 F0IIIn	\\
13305	&	 F4III	\\
20121	&	 F7III+A0V	\\
23010	&	 F5II	\\
34658	&	 F5II	\\
37788	&	 F0IV	\\
56986	&	 F2IV	\\
57852	&	 F0-2IV-V	\\
65925	&	 F5III	\\
72779	&	 G0III	\\
77601	&	 F6II-III	\\
91942	&	 K3-4II	\\
96202	&	 F3IV	\\
104827	&	 F0IV-V	\\
109085	&	 F2III-IV	\\
126251	&	 F4III	\\
127739	&	 F2IV	\\
144183	&	 F2II	\\
164259	&	 F2IV	\\
171237	&	 F3II-III	\\
186357	&	 F1III	\\

187038	&	 K3III	\\
194943	&	 F2IV	\\
200723	&	 F3IV	\\
203784	&	 F6II-III	\\
203842	&	 F5III	\\
207958	&	 F1III	\\
208177	&	 F5IV	\\
219571	&	 F1III	\\
\hline
\end{longtable}
}

\end{document}